 \documentclass[final,5p,times,twocolumn,authoryear]{elsarticle}
\usepackage{amssymb}
\usepackage{lipsum}
\usepackage{natbib}
\usepackage{graphicx}
\usepackage{epsfig}
\usepackage{amsfonts}
\usepackage{amsmath}
\usepackage{amssymb}
\usepackage{amsthm}% in order to use a black box at the end of proofs

\usepackage{color}
\usepackage[colorlinks=true,linkcolor=blue,citecolor=magenta,urlcolor=blue]{hyperref}
\usepackage[utf8]{inputenc}
\usepackage[english]{babel}
\usepackage[T1]{fontenc}
\usepackage{CJKutf8}
\usepackage{multirow}
\usepackage[dvipsnames]{xcolor}
\newcommand{\unit}{1\!\!1}

\usepackage{tikz}

\usepackage{easybmat}
\usepackage{braket}
\usepackage{mathtools}
\usepackage{bbm}
\allowdisplaybreaks
\newcommand{\ketbra}[2]{|#2\rangle\langle#1|}

\newcommand{\be}{\begin{equation}}
\newcommand{\ee}{\end{equation}}
\newcommand{\ba}{\begin{eqnarray}}
\newcommand{\ea}{\end{eqnarray}}

\newcommand{\etal}{{\it{et. al. }}}

\journal{Annals of Physics}

\begin{document}

\begin{frontmatter}

%% Title, authors and addresses

%% use the tnoteref command within \title for footnotes;
%% use the tnotetext command for theassociated footnote;
%% use the fnref command within \author or \affiliation for footnotes;
%% use the fntext command for theassociated footnote;
%% use the corref command within \author for corresponding author footnotes;
%% use the cortext command for theassociated footnote;
%% use the ead command for the email address,
%% and the form \ead[url] for the home page:
%% \title{Title\tnoteref{label1}}
%% \tnotetext[label1]{}
%% \author{Name\corref{cor1}\fnref{label2}}
%% \ead{email address}
%% \ead[url]{home page}
%% \fntext[label2]{}
%% \cortext[cor1]{}
%% \affiliation{organization={},
%%            addressline={}, 
%%            city={},
%%            postcode={}, 
%%            state={},
%%            country={}}
%% \fntext[label3]{}

%% use optional labels to link authors explicitly to addresses:
%% \author[label1,label2]{<author name>}
%% \address[label1]{<address>}
%% \address[label2]{<address>}
%% Author name

%% Author affiliation

\title{Comment on Al-Qasimi. ``Contextuality and quantum discord" [Physics Letters A 449 (2022) 128347]}

%% use optional labels to link authors explicitly to addresses:
%% \author[label1,label2]{}
%% \affiliation[label1]{organization={},
%%             addressline={},
%%             city={},
%%             postcode={},
%%             state={},
%%             country={}}
%%
%% \affiliation[label2]{organization={},
%%             addressline={},
%%             city={},
%%             postcode={},
%%             state={},
%%             country={}}

\author[Jebarathinam]{Chellasamy Jebarathinam}

\affiliation[first]{organization={Physics Division, National Center for Theoretical Sciences},%Department and Organization
            addressline={National Taiwan University}, 
            city={Taipei},
            postcode={106319}, 
            state={},
            country={Taiwan}}

\begin{abstract}
In a paper, Al-Qasimi [\href{https://doi.org/10.1016/j.physleta.2022.128347}{Physics Letters A 449 (2022) 128347}] proposed a criterion for contextuality of two-qubit systems extending Peres' proof of contextuality [\href{https://doi.org/10.1016/0375-9601(90)90172-K}{Phys. Lett. A 151 (1990) 107}]. Using this criterion, Al-Qasimi argued that certain discordant states can be contextual even if they are not entangled. Here I demonstrate that Al-Qasimi's criterion does not, in general, imply contextuality of two-qubit system. However, I point out that Al-Qasimi's criterion is related to a different notion of operational nonclassicality of discordant feature, which is inequivalent to contextuality. This nonclassicality beyond contextuality has been formalized in C. Jebarathinam and R. Srikanth [\href{https://doi.org/10.1142/S0219749925500376}{Int. J. Quantum Inf. 23 (2025)
2550037}]. Identifying Al-Qasimi's criterion as demonstration of quantum correlation beyond contextuality, it then follows that Al-Qasimi's main conclusion that one cannot equate quantum correlation and discordant feature is made precise.
\end{abstract}

%%Graphical abstract
%\begin{graphicalabstract}
%\includegraphics{grabs}
%\end{graphicalabstract}

%%Research highlights
%\begin{highlights}
%\item Research highlight 1
%\item Research highlight 2
%\end{highlights}

\begin{keyword}
%% keywords here, in the form: keyword \sep keyword, up to a maximum of 6 keywords
contextuality \sep quantum correlation \sep quantum discord 

%% PACS codes here, in the form: \PACS code \sep code

%% MSC codes here, in the form: \MSC code \sep code
%% or \MSC[2008] code \sep code (2000 is the default)

\end{keyword}

\end{frontmatter}

%\tableofcontents

%% \linenumbers

%% main text

%\section*{Introduction}
%\label{introduction}

Selftesting of quantum states and measurements is a quantum certification method~\cite{MY04,SBq20} that exploits Bell nonlocality~\cite{Bel64}. Subsequently, selftesting that exploits quantum contextuality (Kochen-Specker contextuality)~\cite{KS67} has been explored~\cite{BRV+19,BCG+22}. The selftesting property exploiting contextuality demonstrates that a quantum certification independently of the Hilbert-space dimension of the quantum system used to observe the phenomenon can be achieved.   

 %consider the contextuality argument of Peres~\cite{Per90} that demonstrates contextuality of two-qubit system in a state-dependent fashion. 
%The argument of Peres uses maximally entangled state. 
%This argument can be studied 
%consider the Peres-Mermin proof of contextuality in a state-dependent fashion~\cite{Per90,Mer90}.

\begin{figure}[t!]
\begin{center}
\includegraphics[width=5.5cm]{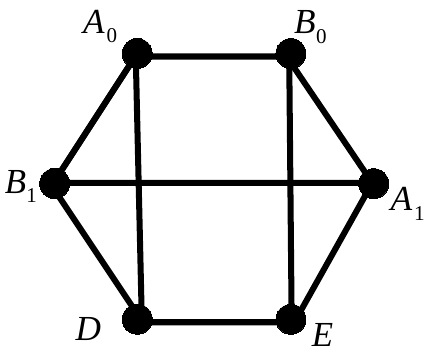}
\end{center}
\caption{Compatibility graph associated with the contextuality scenario associated with the Peres-Mermin proof of contextuality~\cite{Per90,Mer90}. 
\label{Fig:compatibility1}}
\end{figure}

To give an example of selftesting using contextuality, 
consider the contextuality scenario associated with the Peres-Mermin proof of contextuality~\cite{Per90,Mer90}. For this scenario which has five contexts denoted by the following sets of jointly measurable dichotomic observables: $\{A_0,B_0\}$, $\{A_1,B_1\}$, $\{D,E\}$, $\{A_0,B_1,D\}$ and  $\{A_1,B_0, E\}$,  the compatibility graph is depicted in Fig.~\ref{Fig:compatibility1}. 
For the experimental demonstration of contextuality in the aforementioned scenario, Cabello \etal \cite{CFR+08} introduced the following noncontextuality inequality:
\begin{align}\label{dincI4}
\braket{A_0B_0} &+ \braket{A_0B_1D}+\braket{A_1B_0E}+\braket{A_1B_1}  \nonumber \\ 
&- \braket{D E} \le 3,
\end{align}
where each term in the left-hand-side (LHS) is an expectation value of the joint measurement of the observables in each context. Let us choose the observables in the above inequality as follows:
\begin{align}\label{ObsP}
\begin{split}
&A_0=Z_2 \otimes \unit_2, \quad B_0= \unit_2 \otimes Z_2,  \\
&B_1= \unit_2 \otimes X_2, \quad A_1=X_2 \otimes \unit_2, \\
&D=Z_2 \otimes X_2, \quad E=  X_2  \otimes  Z_2.
\end{split}
\end{align}
Here $\unit_2$, $X_2$, $Y_2$ and $Z_2$ denote the identity operator acting on qubit Hilbert space and the Pauli operators given by
$X_2=\ket{0}\bra{1}+\ket{1}\bra{0}$, $Y_2=-i \ket{0}\bra{1}+i\ket{1}\bra{0}$ and 
$Z_2=\ketbra{0}{0}-\ketbra{1}{1}$, respectively.
Measurements of the observables in Eq.~(\ref{ObsP}) 
in the contexts of the scenario on the two-qubit maximally entangled state given by
\be
\ket{\phi^+}=\frac{1}{\sqrt{2}}(\ket{00}+\ket{11}),
\ee
lead to observe LHS of the inequality~(\ref{dincI4}) to take value $5$, which is the maximal quantum violation. Observing the maximal quantum violation of the inequality~(\ref{dincI4}) can be used to selftest the two-qubit quantum system in the above-mentioned state and measurements~\cite{IMO+20,JSS+23}. This way of certification of the quantum state and measurements does not assume the Hilbert-space dimension of the quantum system used to observe the violation of the noncontextuality inequality.

In~\cite{JS25}, in the contextuality scenario of Fig.~\ref{Fig:compatibility1}, nonclassicality beyond contextuality has been formalized. We call this nonclassicality \textit{dimensionally restricted contextuality} (called semi-device-independent contextuality in~\cite{JS25}) for the following reason. In contrast to Kochen-Specker contextuality, demonstration of the nonclassicality beyond contextuality requires that the Hilbert-space dimension of the quantum system is restricted. All contextual correlations have dimensionally restricted contextuality, at the same time, dimensionally restricted contextuality also occurs for correlations that have a noncontextual-hidden-variable model.
Noncontextual correlations that have dimensionally restricted contextuality have supernoncontextuality, which is the contextuality analog of superlocality~\cite{DW15,JAS17} and has been studied in~\cite{JS25}.  

To witness dimensionally restricted contextuality of supernoncontextual correlations, in~\cite{JS25}, the following criterion was introduced using a nonlinear witness of superlocality~\cite{JD23,Jeb25}. This witness was defined in terms of the covariance,  $\texttt{cov}(A_x,B_y)$, of the observables $A_x$ and $B_y$ given by
 \be
 \texttt{cov}(A_x,B_y)=\braket{A_xB_y} -\braket{A_x}\braket{B_y},
 \ee
 where $\braket{A_xB_y}$ and $\braket{A_x}$, and $\braket{B_y}$
 are joint and marginal expectation values, respectively.
The witness of superlocality in terms of the covariances was then given by
\begin{align}
	Q&=\left|\begin{array}{cc}\texttt{cov}(A_0,B_0) & \texttt{cov}(A_1,B_0)\\ 
		\texttt{cov}(A_0,B_1) & \texttt{cov}(A_1,B_1) \end{array}\right|. \label{QC}
\end{align}
Using this witness, the criterion of dimensionally restricted contextuality was given as follows.
Suppose we have a quantum state in $\mathbb{C}^{4}$
and a set of measurements of the contexts in the scenario as in Fig.~\ref{Fig:compatibility1}.
Then, dimensionally restricted contextuality is demonstrated if the witness of superlocality $Q$ in Eq. (\ref{QC}) is nonzero and $\braket{A_0B_1D}=\braket{A_1B_0E}=1$, with  $cov(D,E)>0$. 

To illustrate the main idea of dimensionally restricted contextuality of two-qubit systems, let us consider two-qubit Werner state given by
\begin{align}\label{WerFam}
\rho_W=W \ketbra{\phi^{+}}{\phi^{+}}+ (1-W) \frac{\unit_4}{4},
\end{align}
where $0 \le W\le 1$. $\rho_W$ is entangled for $W>1/3$ (as can be checked via the positive partial transpose criterion~\cite{Per96}), on the other hand, it is discordant, i.e., it has a nonzero quantum discord for any $W>0$~\cite{OZ01,HV01}.  
For the observables given by Eq.~(\ref{ObsP}), the correlation arising from $\rho_W$ gives rise to the LHS of the noncontextuality inequality (\ref{dincI4}) to take value $3W+2$, which implies that $\rho_W$ gives rise to the violation of the noncontextuality inequality for $W>1/3$. The aforementioned correlation arising from $\rho_W$ has a noncontextual-hidden-variable model for $W \le 1/3$, as illustrated in~\cite{JS25}, at the same time, the correlation has dimensionally restricted contextuality for any $W>0$, since it satisfies the criterion for dimensionally restricted contextuality as follows. The correlation gives rise to $Q$ in Eq.~(\ref{QC}) to take valuse as $Q=W^2$ and  $\braket{A_0B_1D}=\braket{A_1B_0E}=1$, with  $\texttt{cov}(D,E)=-W$. Note that if the dimension of the system is not restricted, the above-mentioned correlation arising from $\rho_W$ cannot be taken as nonclassical correlation for $0 < W \le 1/3$. This is because the correlation can be reproduced by a non-discordant state for $0 < W \le 1/3$ if the dimension is not restricted.  

%\section*{Discussion}
To compare the approaches of contextuality and quantum discord
to capture quantum correlations, in~\cite{ALQ22},  extending Peres' proof of contextuality~\cite{Per90}, Al-Qasimi proposed the following criterion for contextuality of two-qubit systems. Define the quantity 
\be \label{ALQ}
\texttt{C}:=|\braket{A_0B_0}\braket{A_1B_1}-\braket{A_0B_1}\braket{A_1B_0}|.
\ee
Al-Qasimi's criterion for contextuality reads as $\texttt{C}>0$ for a suitable choice of observables that imply Peres' proof of contextuality~\cite{Per90}, such as in Eq.~(\ref{ObsP}). Using the above criterion of contextuality to specific classes of two-qubit states, Al-Qasimi claimed that certain discordant states, i.e., states with a nonzero discord can be contextual even if they are not entangled and made the following main conclusion: one cannot equate contextuality to being discordant. In the following, I comment on these claims by Al-Qasimi. 

For the two-qubit Werner state~(\ref{WerFam}),  the above-mentioned Al-Qasimi's criterion for contextuality gives $\texttt{C}=W^2>0$ for any $W>0$. Due to this, Al-Qasimi argued that for Werner states, only when discord is 
zero, the system is noncontextual. However, this contradicts the fact that, in the scenario of Peres' proof of contextuality, the Werner state is noncontextual for $W\le 1/3$, as reviewed before. Moreover, in~\cite{JS25}, it has been shown that for any two-qubit state,  entanglement is required to demonstrate contextuality in the scenario of Peres' proof of contextuality. It then follows that if any discordant state is not entangled, then it cannot be used to imply contextuality in the scenario considered.  Thus, Al-Qasimi's claim that $\texttt{C}>0$ for a suitable choice of observables that imply Peres' proof of contextuality~\cite{Per90} captures contextuality is incorrect. 

However, next I point out that Al-Qasimi's criterion for quantum correlation related to a notion of quantum correlation beyond contextuality formalized in~\cite{JS25}. To this end, let us compare the criterion for dimensionally restricted contextuality using $Q$ in Eq.~(\ref{QC}) and Al-Qasimi's criterion for quantum correlation with Eq.~(\ref{ALQ}). This comparison reveals that Al-Qasimi's criterion for quantum correlation refers to the dimensionally restricted contextuality rather than the standard contextuality of Kochen and Specker.

Finally, note that Al-Qasimi made the following main conclusion: \textit{we cannot equate contextuality to being discordant}. From my discussion above, it follows that the above-mentioned main conclusion of \cite{ALQ22} should be made precise as follows: \textit{we cannot equate dimensionally restricted  contextuality to being discordant}. The fact that dimensionally restricted contextuality in the scenario of Fig.~\ref{Fig:compatibility1} is inequivalent to discordant feature has also been addressed in~\cite{JS25}. I also wish to note that the question of which discordant states have quantum correlations in an operational form has been addressed in~\cite{JDK+25,JKC+25}.

\textit{Acknowledgement.–} This work was supported by the National Science and Technology Council, the Ministry of Education (Higher Education Sprout Project NTU-113L104022-1), and the National Center for Theoretical Sciences of Taiwan.

\bibliographystyle{elsarticle-harv} 
\bibliography{example}

\end{document}